# Multimodal Drivers’ Emotion Recognition and Safety-Oriented Intervention for Intelligent Transportation Systems

Chang Liu[1,3†], Dalai Mengke[2†], Hanbo Zhou[1†], Jia Hu[4], Péter Mihajlik[2], Tamás Szirányi[1,3]

***Abstract*— Driver emotions can affect risk perception, decision-making, and vehicle control under complex road conditions. Existing studies mainly focus on driver emotion recognition, while limited attention has been given to context-aware intervention that jointly considers driver emotion and road perception. This paper proposes a safety-prioritized multimodal driver assistance framework that analyzes speech-derived emotional cues and visual road conditions to generate structured driving interventions. The framework first provides road safety reminders and then generates emotion-aligned verbal support. We construct a multimodal dataset by aligning emotional speech signals with structured road environment descriptors and introduce the CARE (Context-Aware Road–Emotion Evaluation) score to jointly evaluate emotion recognition, risk identification, and intervention generation. Experimental results show that the proposed framework balances environmental risk reporting and emotion-aware verbal regulation, providing a feasible safety-driven direction for intelligent transportation systems.**

## I. Introduction

Road safety is jointly determined by environmental conditions, vehicle dynamics, and human cognitive and affective states. Among these factors, driver emotion plays a critical yet often overlooked role in shaping perception accuracy, decision-making efficiency, and hazard response under both routine and emergency driving scenarios [1]. Empirical studies indicate that negative or high-arousal emotional states significantly increase accident risk, while stable or neutral affect contributes to improved driving performance [2]. Importantly, this influence persists even in vehicles equipped with advanced driver-assistance systems (ADAS) or partial automation, where emotional fluctuations can alter takeover time, attention allocation, and reaction reliability.

Existing research on driver emotion analysis predominantly focuses on recognition accuracy using visual, acoustic, or physiological signals [3], [4]. In driving-related scenarios, most approaches rely on facial expression modeling and speech-based emotion classification [5]. Although these methods achieve improved recognition performance, they largely treat emotion detection as a standalone task and rarely explore how detected emotional states can be leveraged for real-time safety enhancement or context-aware intervention. As a result, there remains a significant gap between emotion perception and actionable safety support.

Beyond recognition, integrating driver emotion with dynamic road context to enable proactive intervention remains underexplored. A practical safety-oriented system should not only detect emotional states but also jointly model environmental risk cues and generate structured, controllable responses that prioritize hazard mitigation before emotional regulation. However, existing multimodal systems typically lack this safety-first generation mechanism and do not explicitly model the interaction between emotional instability and road risk exposure.

To address this gap, we propose a safety-prioritized multimodal framework that jointly models speech-derived emotional cues and visual road conditions to enable structured driving intervention. Instead of merely predicting emotions, our system generates sequential outputs consisting of (1) context-aware road safety reports and (2) emotion-aligned verbal support. This design ensures that risk awareness is prioritized before affective intervention, reducing distraction while preserving emotional stability.

To support this framework, we construct a multimodal dataset that integrates emotional speech signals with structured road environment descriptors and corresponding intervention annotations. Furthermore, we introduce our CARE score (Context-Aware Road–Emotion Evaluation), a composite metric designed to jointly evaluate emotion recognition reliability, risk identification accuracy, and safety-consistent response generation.

In summary, this work makes the following contributions:

- **A Safety-First Multimodal Intervention Paradigm:** We formulate a structured intervention framework that prioritizes road-risk reasoning followed by targeted driver emotional support, bridging the gap between emotion recognition and actionable safety assistance.
- **Multimodal Dataset Construction:** We build a dataset combining emotional speech, visual road conditions, and intervention annotations to enable research on context-aware emotional comfort and safety-driven assistance.
- **Joint Evaluation Mechanism:** We propose the CARE score to comprehensively measure emotion classification performance, risk detection accuracy, and semantic quality of generated interventions under a unified safety-oriented metric.

To our knowledge, this work represents an early attempt to

[1]Department of Networked Systems and Services, Faculty of Electrical Engineering and Informatics, Budapest University of Technology and Economics (BME), Budapest, Hungary.

[2]Department of Telecommunications and Artificial Intelligence, Faculty of Electrical Engineering and Informatics, Budapest University of Technology and Economics (BME), Budapest, Hungary.

[3]Machine Perception Research Laboratory, HUN-REN Institute for Computer Science and Control (HUN-REN SZTAKI), Budapest, Hungary.

[4]Key Laboratory of Road and Traffic Engineering, Ministry of Education, Tongji University, Shanghai 201804, China.

[†]Chang Liu, Dalai Mengke, and Hanbo Zhou contributed equally to this work.[†]Equal contribution.

Corresponding authors: Dalai Mengke and Chang Liu. Email: kedalai.meng@edu.bme.hu; changliu@hit.bme.hu.

integrate structured road-risk reporting and emotion regulation within a unified multimodal safety intervention framework for intelligent transportation systems.

## II. Related Study

Driver emotion analysis has become an important research direction for improving driving safety and enhancing human–vehicle interaction. Existing studies primarily exploit visual, acoustic, and physiological signals, as well as their multimodal fusion, to improve recognition robustness and accuracy. For facial-expression-based recognition, Roka and Rawat [6] fine-tuned a Vision Transformer on AffectNet, while Manavand et al.[7] introduced an ILAB-CNN with modified Squeeze-and-Excitation blocks, achieving competitive performance on public benchmarks. In the acoustic domain, Xie et al.[8] combined blind source separation with a ResNet34 backbone to mitigate in-cabin noise, demonstrating improved speech emotion recognition performance. Cross-modal approaches further enhance robustness: Na Ying et al.[9] integrated speech and facial features via attention-based fusion, and Guoliang et al.[10] leveraged a dual-path Transformer to fuse facial expressions and rPPG signals.

Although these approaches achieve strong recognition accuracy, they primarily optimize classification performance under offline benchmarks and rarely extend to safety-oriented reasoning or real-time intervention mechanisms. Most existing systems treat emotion recognition as an isolated perception task without explicitly connecting emotional states to structured driving behavior or interaction strategies.

Beyond emotion recognition, recent research explores emotion regulation and adaptive interaction within vehicle environments. Braun et al.[11] showed that empathic feedback is generally preferred over neutral responses, while Choe and Jeon [12] demonstrated that empathy-based voice agents can improve driving stability and reduce negative affect. Multimodal environmental cues such as ambient lighting, music, and vibration have also been shown to influence driver emotional states. With the emergence of large language models (LLMs), interactive in-vehicle assistance systems have gained new capabilities. GPT-based dialogue systems have been reported to enhance lane-keeping performance and driver trust [13], and emotionally aligned prompt strategies have further improved safety-related driving behavior [14]. Recent multimodal driving studies have extended this direction by incorporating audio signals into vision–language–action (VLA) models. For instance, Guo et al.[15] introduced an audio-conditioned VLA framework in which speech cues are used to influence trajectory planning and vehicle control decisions. Such approaches focus on action-level adaptation, enabling motion modulation according to user instructions or emotional tone.

However, existing regulation-oriented and VLA-based approaches typically emphasize either emotion perception or trajectory-level control adaptation, without explicitly integrating structured real-time road perception with emotion-aware semantic safety reporting. The absence of a unified framework that jointly reasons about environmental risk and driver emotional state at the interaction level limits the development of context-aware safety mediation systems. To address this gap, we propose an integrated multimodal framework that jointly models speech-derived emotional cues and visual road conditions, and leverages large language models to generate structured safety reports followed by emotion-aligned verbal support.

## III. Methodology

This section presents the proposed safety-prioritized multimodal intervention framework, designed to integrate driver emotional states with road environment context to generate timely safety guidance and emotion-aware verbal support.

As illustrated in Fig. 1, the system consists of four major modules: (1) **multimodal data construction**, for preparing paired image and speech samples with structured annotations; (2) **feature extraction and cross-modal representation learning**, which extracts modality-specific features and fuses them through stacked co-attention layers; (3) **large language model-based structured intervention generation**, which produces prioritized safety reports and emotion-aligned guidance; and (4) **composite safety-aware evaluation**, which quantitatively assesses model performance.

Given aligned speech and road image inputs, modality-specific features are first extracted using pretrained encoders. These features are then projected into a shared latent space and fused through co-attention layers, enabling the model to capture interactions between driver emotions and road conditions. The resulting fused representation is transformed into structured prompts and fed into a large language model to generate prioritized road safety reports, followed by emotion-aligned verbal support.

To quantitatively assess system performance, we introduce the **CARE score (Context-Aware Road–Emotion Evaluation)**, which jointly evaluates emotion recognition reliability, risk detection accuracy, and the semantic quality of generated interventions.

### A. Dataset Construction

Since no public dataset directly supports safety-oriented multimodal intervention in driving scenarios, we construct a customized multimodal dataset by integrating and augmenting existing public datasets to enable joint modeling of driver emotion and structured road context.

**Design Motivation.** The dataset is specifically designed to support structured risk reasoning and emotion-aware intervention generation. Instead of treating visual and speech data independently, we construct aligned multimodal pairs with explicit safety attributes and intervention-oriented annotations to facilitate supervised training of the proposed framework.

**Visual Modality.** We adopt the BDD100K dataset [16], which provides large-scale urban driving scenes with annotations including weather, time of day, and road conditions. To enable structured reasoning over road context, we extract 42 quantitative descriptors covering traffic density, object distribution, occlusion level, visibility estimation, road geometry, and safety-relevant attributes. These descriptors are normalized and transformed into structured textual tokens to serve as explicit supervision signals for downstream training.

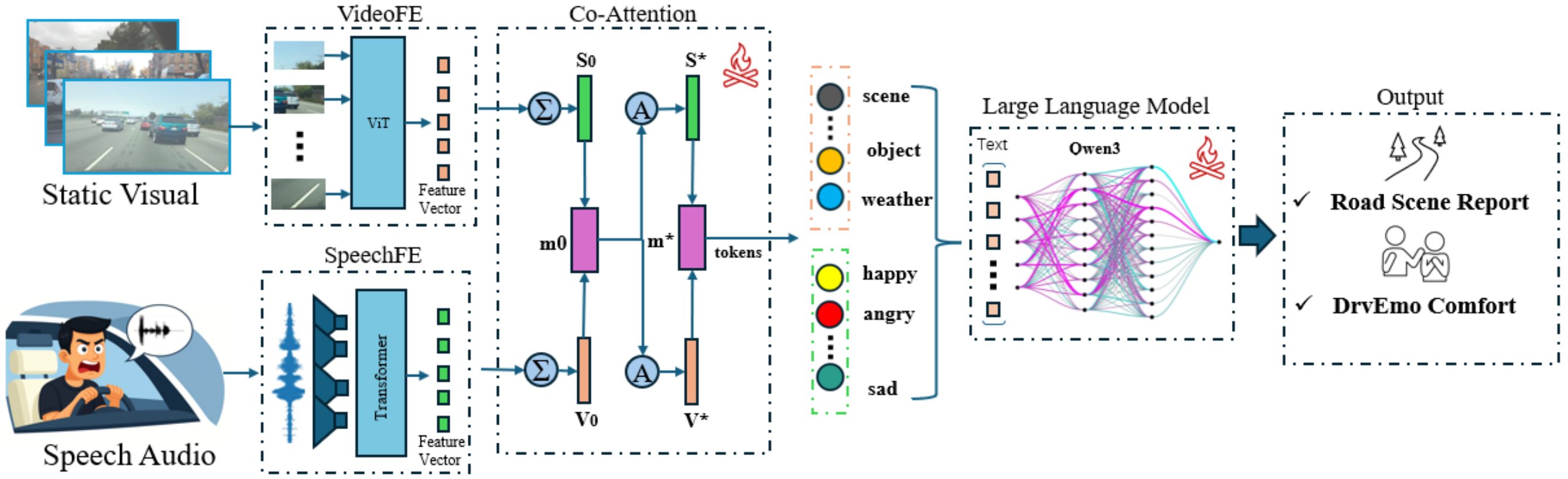

Fig. 1. Multimodal Framework for Emotion-Aware Safety Intervention

**Audio Modality.** For speech emotion modeling, we utilize the CREMA-D dataset [17], which consists of English emotional speech recordings. To simulate realistic in-vehicle acoustic environments, we apply controlled augmentation strategies including time stretching (0.9–1.1× speed perturbation), reverberation, gain adjustment (-6 to +6 dB), SNR perturbation (5–25 dB), codec distortion, spectral masking, and silence insertion (0.1–0.5 s). These augmentations improve robustness under noisy and dynamic driving conditions. All audio samples are standardized to fixed-length clips (e.g., 3 seconds at 16 kHz), and emotion labels are converted into structured textual representations to maintain consistency with the visual descriptors.

**Multimodal Alignment and Annotation.** To construct paired multimodal instances, we perform scale normalization and structured feature alignment between visual descriptors and emotion labels. The aligned features are then fed into a large language model-based annotation pipeline to generate intervention-oriented target texts, including:

- Structured road safety reports,
- Emotion-aware verbal responses,
- Combined safety-emotion intervention instructions.

This automatic annotation process enables scalable generation of structured supervision signals for training the proposed framework.

It should be noted that the constructed multimodal pairs are synthetically aligned rather than naturally synchronized recordings collected from real in-vehicle scenarios. Therefore, the dataset is intended to evaluate framework feasibility under controlled multimodal conditions, rather than to fully model the natural temporal coupling between driver emotion and dynamic road context.

**Dataset Statistics.** The visual modality follows the original dataset split, while the audio modality is divided into 5,896 training clips, 732 validation clips, and 814 testing clips. During training, each speech clip is paired with multiple road-scene images to maximize the utilization of the available visual dataset and increase multimodal diversity. For validation and testing, a one-to-one pairing strategy is adopted to avoid artificial duplication during evaluation. All subsets are separated before multimodal pairing to prevent data leakage.

### *B. Multi-Modal Representation Learning*

To generate safety-oriented interventions, our framework jointly models visual road context and speech-derived emotional cues through a structured three-stage pipeline: feature extraction, cross-modal fusion, and token sequence generation for large language model input.

**Visual Encoder.** Road images of size $[B, 3, 224, 224]$ are processed by **ViT-Base-Patch16** [18], producing 196 patch embeddings of 768 dimensions each (excluding the CLS token). The encoder consists of 12 transformer layers, and its parameters are frozen to preserve pre-trained visual representations.

**Audio Encoder.** Speech signals of length $T_a$ are encoded using a pretrained Wav2Vec2 model [19]. In our implementation, we initialize the encoder from `facebook/wav2vec2-large-xlsr-53` and further fine-tune it on English speech emotion data. The encoder consists of 24 Transformer layers with 1024-dimensional contextual embeddings. During multimodal training, the audio encoder parameters are frozen to preserve pretrained acoustic representations.

**Motivation for Cross-Modal Fusion.** Unlike simple concatenation or late fusion strategies, safety-critical driving scenarios require fine-grained alignment between transient emotional states and dynamic road conditions. To address this, we adopt a stacked Co-Attention mechanism that explicitly models bidirectional dependencies across modalities, allowing the system to prioritize features relevant to both emotional and environmental risks.

**Co-Attention Fusion.** Visual and audio embeddings are first projected into a shared 512-dimensional latent space. A 2-layer Co-Attention module is applied, where each layer includes:

1. *Intra-modal self-attention* to strengthen modality-specific

dependencies. 2. *Bidirectional cross-attention* to integrate visual and acoustic information:

$$\tilde{\mathbf{V}} = \mathrm{MHA}(Q = \mathbf{V}, K = \mathbf{A}, V = \mathbf{A}), \quad (1)$$

$$\tilde{\mathbf{A}} = \mathrm{MHA}(Q = \mathbf{A}, K = \mathbf{V}, V = \mathbf{V}), \quad (2)$$

where $\mathbf{V}$ and $\mathbf{A}$ denote the visual and audio token sequences, respectively, and tildes represent cross-modally enhanced embeddings.

The co-attention outputs are processed through feed-forward networks with residual connections. A global average pooling followed by concatenation produces a fused vector of size $[B, 1024]$, which is projected via two-layer MLP fusion to $[B, 512]$.

**Multimodal Token Generation.** The fused representation is expanded using a sequence generator into a multimodal token sequence of $[B, 32, 4096]$, with learnable positional encodings added to provide position-aware token identities. These tokens serve as structured context prompts for the downstream large language model, guiding safety-prioritized intervention generation.

**Computational Considerations.** The co-attention design ensures a memory-efficient interaction between modalities with complexity $O(N_v \cdot N_a)$, where $N_v$ and $N_a$ are the numbers of visual and audio tokens. This enables fine-grained cross-modal alignment without excessive computational cost, suitable for prototype-level near-real-time evaluation.

### C. LLM-Based Structured Intervention Generation

To translate fused multimodal representations into safety-aware structured interventions, the token sequence produced by the fusion module is prepended as a prefix prompt to the Qwen3-8B large language model [20], which contains 36 transformer layers. Instead of fully fine-tuning the large language model, we adopt parameter-efficient adaptation to preserve pretrained reasoning capability while enabling task-specific intervention generation.

**Parameter-Efficient Adaptation.** We apply Low-Rank Adaptation (LoRA) to five key projection matrices in each transformer layer: `q_proj`, `v_proj`, `gate_proj`, `up_proj`, and `down_proj`. The LoRA rank is set to $r = 16$, introducing trainable low-rank matrices $A \in \mathbb{R}^{d_{in} \times r}$ and $B \in \mathbb{R}^{r \times d_{out}}$ for each adapted projection. This design reduces the trainable parameter cost while preserving expressive transformation capability for safety-oriented multimodal reasoning. The LoRA adaptation introduces approximately 39M trainable parameters. Together with the co-attention projectors and the sequence generator, the total trainable parameter budget is approximately 106M. The original weights of the Qwen3-8B model, as well as the ViT encoder and Wav2Vec2-Large encoder, remain frozen to prevent catastrophic forgetting and preserve general multimodal understanding capability.

**Intervention Generation Mechanism.** The fused multimodal tokens act as structured contextual prefixes that condition autoregressive generation. The model outputs prioritized road safety reports followed by emotion-aligned verbal interventions, ensuring a safety-first generation order.

### D. Training Strategy

**Optimization Design.** We adopt a multi-level learning rate strategy to stabilize parameter adaptation across heterogeneous modules:

- **LoRA parameters**: relatively high learning rate to accelerate task-specific adaptation.
- **Cross-modal projection layers  sequence generator**: moderate learning rate to stabilize multimodal alignment.
- **Token embeddings and sequence generator**: low learning rate to preserve pretrained language priors.

This heterogeneous optimization strategy balances convergence stability and representation flexibility in safety-critical multimodal reasoning.

**Training Objective.** The model is trained using masked autoregressive cross-entropy loss over generated intervention tokens. Given target sequence $\mathbf{y}$ and predicted logits $\hat{\mathbf{y}}$, supervision is applied only to intervention-related tokens:

$$\mathcal{L} = -\sum_{t \in \mathcal{M}} \log P(y_t | y_{<t}, \mathbf{x}_{vision}, \mathbf{x}_{audio}), \quad (3)$$

where $\mathcal{M}$ denotes masked positions corresponding to structured safety reports and emotion-aware responses.

By masking conditioning prompt tokens, the objective avoids computing training loss on input prompt positions and encourages controlled structured generation.

### E. Composite Safety-Aware Evaluation

To provide a unified evaluation of multimodal driver monitoring performance, we define a composite safety-aware score, denoted as CARE. The score integrates three complementary aspects of the task: driver emotion recognition, driving risk factor detection, and the semantic quality of the generated intervention tip.

For driver emotion recognition, we compute the emotion accuracy as

$$A_{\text{emotion}} = \frac{100}{N} \sum_{i=1}^{N} \mathbb{I}(\hat{c}_i = c_i), \quad (4)$$

where $\hat{c}_i$ and $c_i$ denote the predicted and ground-truth emotion categories for sample $i$, respectively, and $\mathbb{I}(\cdot)$ is the indicator function.

For risk factor detection, we compute the sample-level set-based F1 score between the predicted risk factor set $\hat{\mathcal{R}}_i$ and the ground-truth set $\mathcal{R}_i$, and average it over all test samples:

$$F_{\text{risk}} = \frac{100}{N} \sum_{i=1}^{N} \frac{2|\hat{\mathcal{R}}_i \cap \mathcal{R}_i|}{|\hat{\mathcal{R}}_i| + |\mathcal{R}_i|}. \quad (5)$$

For the intervention tip, we evaluate semantic similarity between the generated tip $\hat{t}_i$ and the reference tip $t_i$ using BERTScore. The reference tips are constructed from the corresponding emotion labels and road-risk annotations using safety-oriented intervention templates. The average tip quality score is computed as

$$S_{\text{tip}} = \frac{100}{N} \sum_{i=1}^{N} \mathrm{BERTScore}(\hat{t}_i, t_i). \quad (6)$$

All three components are normalized to the range $[0, 100]$ before aggregation. The final CARE score is then defined as

$$\text{CARE} = 0.45 \cdot A_{\text{emotion}} + 0.45 \cdot F_{\text{risk}} + 0.10 \cdot S_{\text{tip}}. \quad (7)$$

The weights assign equal importance to emotion recognition and risk detection, as both are safety-critical structured prediction tasks. Tip quality is assigned a smaller weight because semantic similarity mainly reflects the linguistic alignment of the generated intervention with the reference response, while safety-critical correctness is primarily captured by the emotion and risk detection metrics. CARE is used only as a unified evaluation score and is not directly optimized during training. BERTScore mainly reflects semantic similarity rather than safety-critical correctness.

## IV. Experimental Results

### A. Implementation Details

Experiments are conducted on a single NVIDIA A100 GPU. The framework is implemented with mixed-precision (bf16) training to improve memory efficiency and computational speed.

**Training Configuration.** The model is trained with a batch size of 4 for a maximum of 5 epochs on the combined multimodal dataset. The best-performing checkpoint is selected at epoch 3 based on validation performance, while later epochs exhibit signs of overfitting on the synthetically aligned multimodal data. The total training time is approximately 15 hours. Optimization is performed using AdamW with the following hyperparameters: learning rate $1 \times 10^{-6}$, LoRA learning rate $1 \times 10^{-4}$, Co-Attention learning rate $5 \times 10^{-5}$, weight decay 0.01, $\beta_1 = 0.9$, $\beta_2 = 0.999$, and $\epsilon = 10^{-8}$.

A separate learning rate is assigned to different modules to stabilize parameter-efficient fine-tuning, enabling fast adaptation of LoRA layers while preserving pretrained representations. Because the visual encoder, audio encoder, and backbone large language model remain frozen during training, only the LoRA layers, co-attention projectors, and sequence generator are optimized.

**Model Configuration.** LoRA is applied to five projection matrices in each transformer layer of Qwen3-8B with rank $r = 16$. This configuration introduces approximately 39M trainable parameters through LoRA adaptation. Together with the co-attention projectors and sequence generator, the total trainable parameter budget is approximately 106M. The visual and audio encoders remain frozen during training. Only the LoRA parameters, Co-Attention projectors, and sequence generator are updated, resulting in an efficient adaptation strategy for safety-oriented multimodal reasoning.

### B. Quantitative Results

Table I reports quantitative comparisons among the proposed framework, single-modality baselines, conventional multimodal fusion strategies, and the ablated variant without the co-attention module. The results show that removing the explicit co-attention interaction leads to slight reductions in emotion accuracy and overall CARE score, suggesting that cross-modal interaction contributes to more balanced multimodal reasoning and improved intervention consistency. Although the numerical improvement over Late Fusion remains moderate, the proposed framework achieves the highest overall CARE score across all evaluated settings.

TABLE I
BASELINE COMPARISON AND QUANTITATIVE EVALUATION RESULTS

| Model | Emotion Acc. | Risk F1 | Tip Quality | CARE |
|---|---|---|---|---|
| Audio Only | 62.5 | 48.2 | 66.43 | 56.46 |
| Vision Only | 59.1 | 83.4 | 65.97 | 70.72 |
| Simple Concat | 61.5 | 81.8 | 66.32 | 71.12 |
| Late Fusion | 62.9 | **83.6** | **67.19** | 72.64 |
| w/o Co-Attention | 62.8 | 80.0 | 65.85 | 70.85 |
| Proposed Model | **63.8** | 83.5 | 66.83 | **72.97** |

The results also reveal modality dominance across different sub-tasks. Speech signals contribute primarily to emotion recognition, while visual information dominates environmental risk perception. Therefore, the main advantage of multimodal fusion lies in improving contextual alignment and safety-oriented intervention coherence rather than substantially increasing isolated classification accuracy. The relatively limited performance gap also indicates that safety-oriented multimodal intervention remains challenging under heterogeneous and weakly aligned multimodal data conditions. Future work will further investigate stronger fusion mechanisms and evaluate the framework on synchronized in-vehicle multimodal datasets.

### C. Qualitative and Efficiency Analysis

Fig. 2 shows representative intervention examples under different emotional and environmental conditions. The system first generates road safety reports and then provides emotion-aware support, reflecting the proposed safety-first intervention design. To evaluate computational efficiency, we measure inference latency on a single NVIDIA A100 GPU. The model achieves approximately 4 FPS (0.25 s per sample), corresponding to a throughput of 349 tokens/s, with 39.5 GB GPU memory consumption. These results indicate near-real-time feasibility on a high-end GPU prototype platform, while embedded in-vehicle deployment still requires future optimization through quantization, distillation, or lightweight architectures.

## V. Conclusion and Future Work

This paper proposed a safety-prioritized multimodal framework for emotion-aware driving assistance. By integrating visual road context and speech-derived emotional cues through co-attention-based representation learning, the system generates structured road safety reports followed by emotion-aligned verbal support. Experimental results show that the proposed framework improves overall CARE performance compared with single-modality and conventional fusion baselines, while maintaining a safety-first intervention order.

Several limitations remain. The current dataset is constructed through domain adaptation and synthetic alignment rather than naturally synchronized in-vehicle recordings,

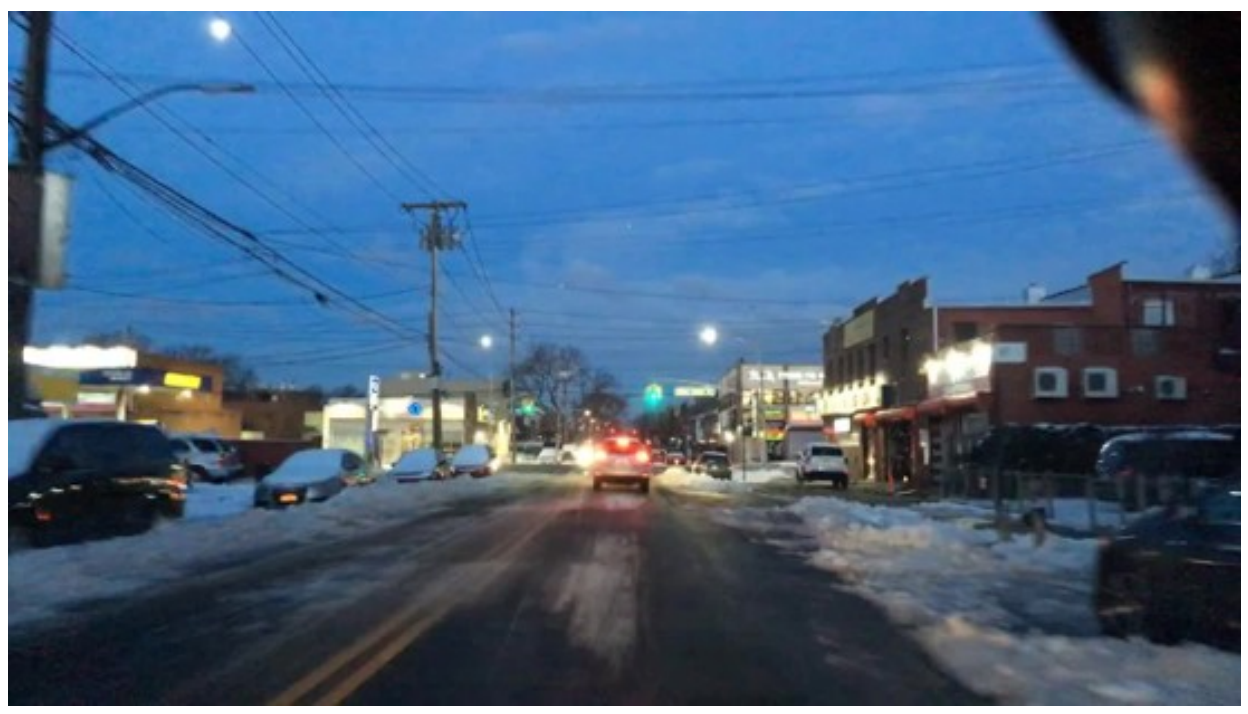

**Example 1 (Sad + Snowy Condition)**
**Risk Factors:** Low visibility, Close front vehicle, Snow
**Road Safety Report:** It is snowy and the vehicle ahead is close. Please brake gently and maintain smooth control.
**Emotion-Aware Support:** Snowy conditions may feel heavier today. It is okay to feel that way; let us proceed calmly and steadily.

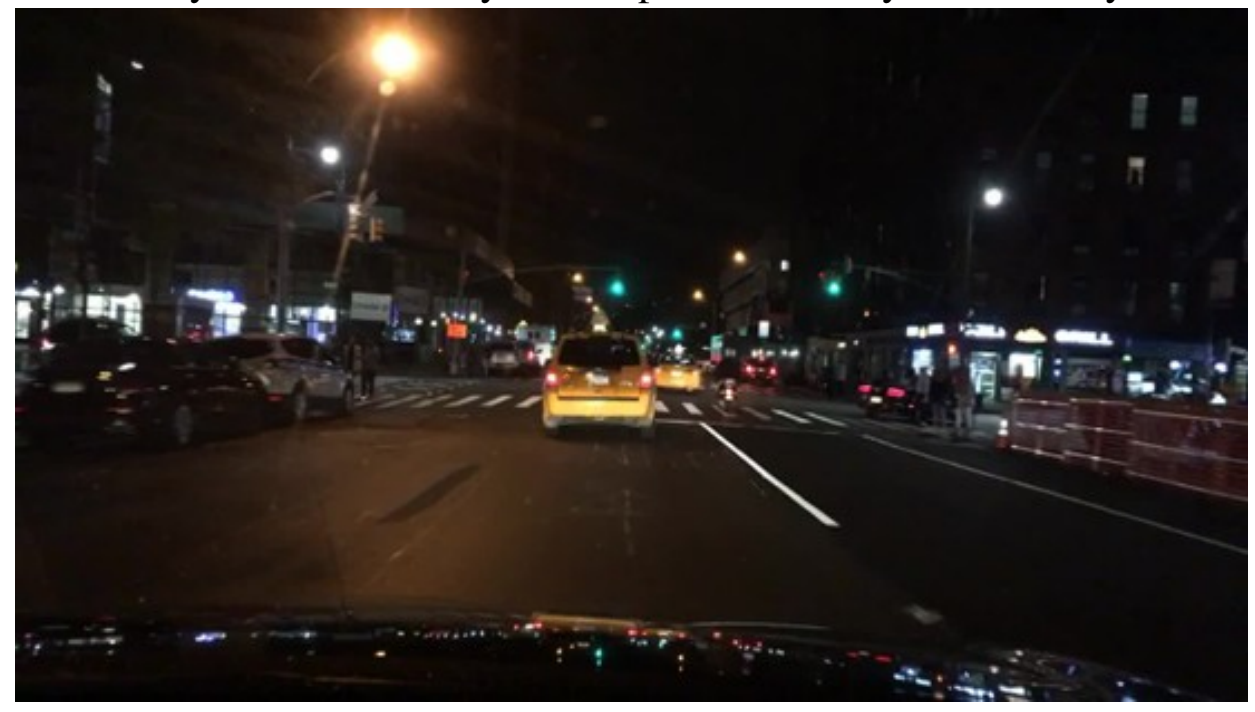

**Example 2 (Fear + Night Driving)**
**Risk Factors:** Night driving, Low visibility, Close front vehicle
**Road Safety Report:** Visibility is reduced at night. Please maintain extra distance and proceed smoothly.
**Emotion-Aware Support:** Low visibility can feel unsettling. You are handling it well; steady control ensures safety.

Fig. 2. Representative examples of safety-oriented intervention generation under different emotional and environmental conditions.

which may limit generalization to real driving environments. In addition, the emotion perception module relies primarily on speech signals, while facial expressions, physiological signals, vehicle speed, and driving behavior dynamics could further improve robustness. Future work will investigate synchronized real-world data collection, lightweight deployment, and richer multimodal sensing for practical intelligent transportation systems.

## ACKNOWLEDGMENTS

The authors would like to thank Prof. Vilmos Simon and the BME-HIT MEDIANETS Laboratory for supporting this research, conference registration, and participation in IEEE ITSC 2026. OTKA K139485 is noted as well. The source code and dataset construction scripts are available at: `https://github.com/tomspter/MultimodalEmotionDrive-ITS`.